\documentclass[fleqn,12pt,twoside]{article}
\usepackage{espcrc1}
\usepackage{graphicx}
\usepackage{epsfig}

\title{Canonical aspects of strangeness enhancement}
\author{A. Tounsi\address{Laboratoire de Physique Th\'eorique et
Hautes Energies, \\
Universit\'e Paris 7, 2 place Jussieu, F-75251 Cedex 05},
        A. Mischke\address{Gesellschaft f\"ur Schwerionenforschung (GSI),
\\ D-64220 Darmstadt, Germany} and
        K. Redlich\address{Fakult\"at f\"ur Physik, Universit\"at Bielefeld,\\
Postfach 100 131, D-33501 Bielefeld, Germany
       }'\address{Institute of Theoretical Physics University of Wroclaw,\\
PL-50204 Wroclaw, Poland  } }
\begin{document}

\maketitle
\begin{abstract}
 Strangeness enhancement (SE) in heavy ion collisions can be understood
 in the statistical model
on the basis of canonical suppression. In this formulation,
SE is a consequence of the transition from
canonical to the asymptotic  grand canonical limit and is
predicted to be a   decreasing function of collision energy. This
model predictions are consistent with the recent NA49 data on $\Lambda$
enhancement at $p_{lab}=40, 80, 158$ GeV. 

\end{abstract}
\section{INTRODUCTION}
  The enhanced production of strange particles
in nucleus-nucleus (AA)  relatively to proton-proton (pp) or
proton-nucleus (pA) collisions, was since long, argued
\cite{rafelski1} to be a signal of a Quark-Gluon Plasma (QGP)
formation. This enhancement is seen in all experiments  at all
energies from AGS to SPS up to RHIC. Furthermore, enhancement of
strange  baryons and anti-baryons was predicted to depend on their
strangeness content and to appear in a typical hierarchy,
$E_{\Lambda}  < E_{\Xi} < E_{\Omega}$, which was observed in
particular by the WA97 \cite{andersen}, NA57 \cite{carrer} and
NA49 \cite{NA49} Collaborations. These enhancement features and
hierarchy, are argued to be expected  at all energies as a
consequence of a  canonical thermal phase-space suppression.
\section{CANONICAL STRANGENESS SUPPRESSION}
When describing a statistical system there are at least two ways
of implementing the conservation laws of quantum numbers. In the
Grand Canonical (GC) formulation conservation of a quantum number
is ensured on average by use  of the corresponding chemical
potential. This description is only valid when particles, with a
given conserved charge,  are produced with a large multiplicity
(system of a large volume and/or high temperature). In the
Canonical formulation (C) conservation of a quantum number is
implemented exactly on an event-by-event basis. This approach is
relevant when particles carrying a given conserved charge are
produced with a small multiplicity
 (system of
small volume and/or low temperature).
In the usual definition of SE one
compares the yield per participant of a given particle $i$ with
strangeness $s$ in the
large AA system to the yield per participant, of the same
particle, in the small pp or pA system. Assuming that
the volume parameter scales with the number of participants,
$V\sim N_{part}$, the enhancement of this particle is given by
\begin{equation}\label{enh}
E_s^i={\left(n_s^i\right)^{AA}\over\left(n_s^i\right)^{pp}}
\end{equation}
where $n_s^i$ is the number density of particle $i$.
 From the above considerations, the pp (or pA) system has to be
described  in the canonical formulation. In principle, all
conserved charges (baryon number $B$, electric charge $Q$, and
strangeness $S$) have to be treated canonically. However, with a
good approximation,  $B$ and $Q$ can be treated grand canonically
even in high energy pp collisions. This results in a slight
overestimate of the canonical effect \cite{keranen}.

The canonical partition function of a gas  of particles and
resonances having strangeness $s=0,\pm 1,\pm 2, \pm 3$, and with
total strangeness $S=0$  can be written as \cite{cleymans}

\begin{equation}\label{ZC}
Z^C_{S=0}(T,V)={1\over {2\pi}}
       \int_{-\pi}^{\pi}
    d\phi~ \exp{\left(\sum_{n=- 3}^3S_ne^{in\phi}\right)}
\end{equation}
where
$S_n=V \sum_k Z_k^1$ , is the sum
 over all particles and resonances  carrying strangeness $n$
and where
$Z_k^1\equiv ({{g_k}/{2\pi^2}}) m_k^2TK_2({m_k/T})
\exp{(B_k\mu_B/T+Q_k\mu_Q/T)}$
is the one-particle partition function of particle $k$ with
 multiplicity $g_k$, mass $m_k$, baryon number $B_k$, electric
charge $Q_k$ and the corresponding chemical potentials $\mu_B$ and
$\mu_Q$.
From Eq.(\ref{ZC}) one can derive \cite{peter}  the
canonical density $(n_s^i)^C$ and show that
\begin{equation}
(n_{s}^i)^C= (n_{s}^i)^{GC}F_s(T,V)
\end{equation}
where $(n_{s}^i)^{GC}$ is the grand canonical density and
$F_s(T,V)$ is the  suppression factor that measures a deviation of
$(n_{s}^i)^C$ from its asymptotic,  grand canonical value. For
large $V$ and/or $T$, $F_s(T,V)\to 1$, and $(n_{s}^i)^C$ reaches
its grand canonical value.
\begin{figure}[htb]
\vskip -1.cm
\begin{minipage}[t]{80mm}
\psfig{width=7.cm, height=14.pc,figure=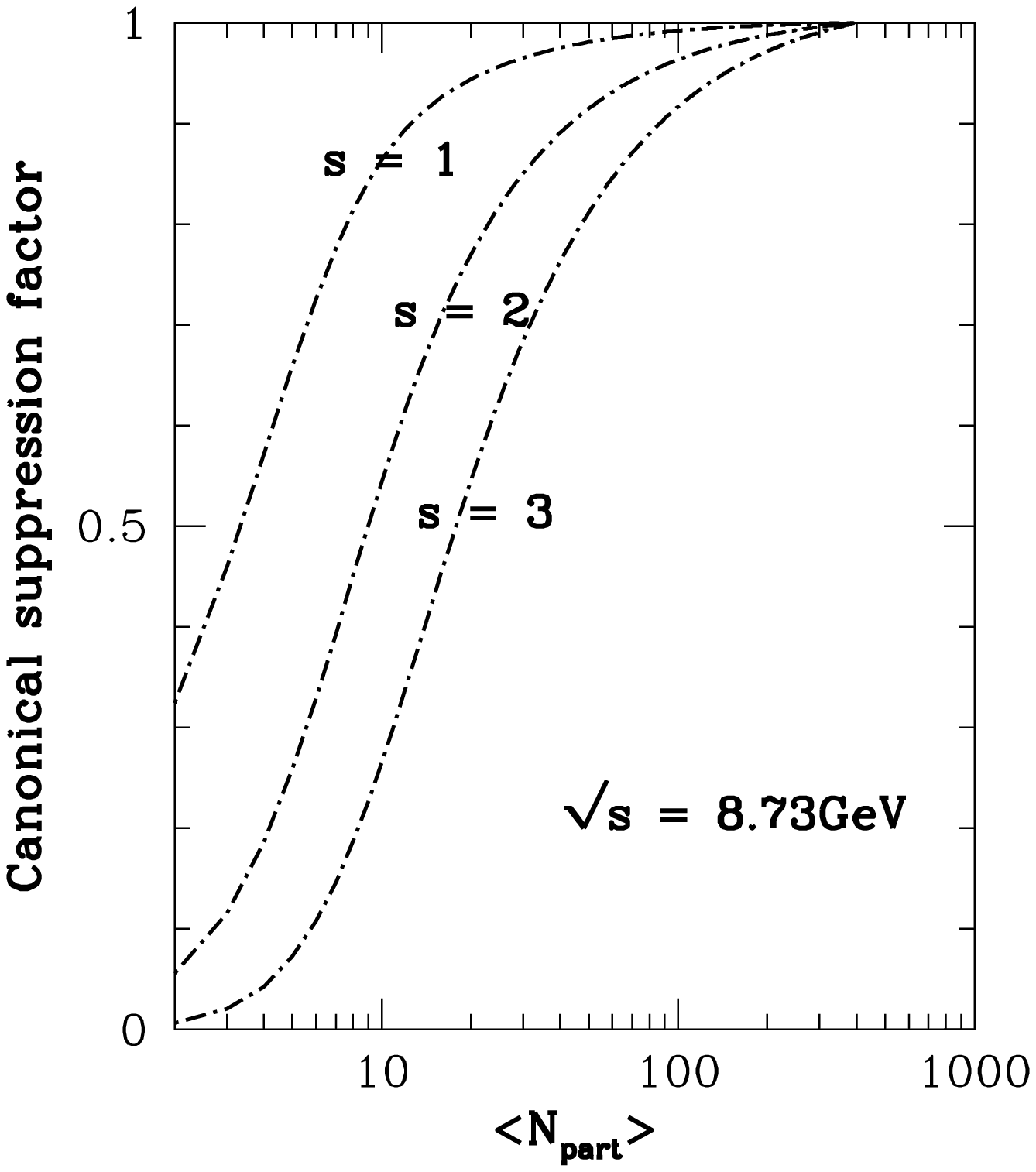} \vskip
-1.cm \caption{Canonical suppression factor for different of
particle strangeness: $s=1,2,3$ at  energy $\sqrt{s}=8.73$ GeV.
\hfill}\label{can}
\end{minipage}
\hspace{\fill}
\begin{minipage}[t]{75mm}
\psfig{width=7.cm,height=14.pc,figure=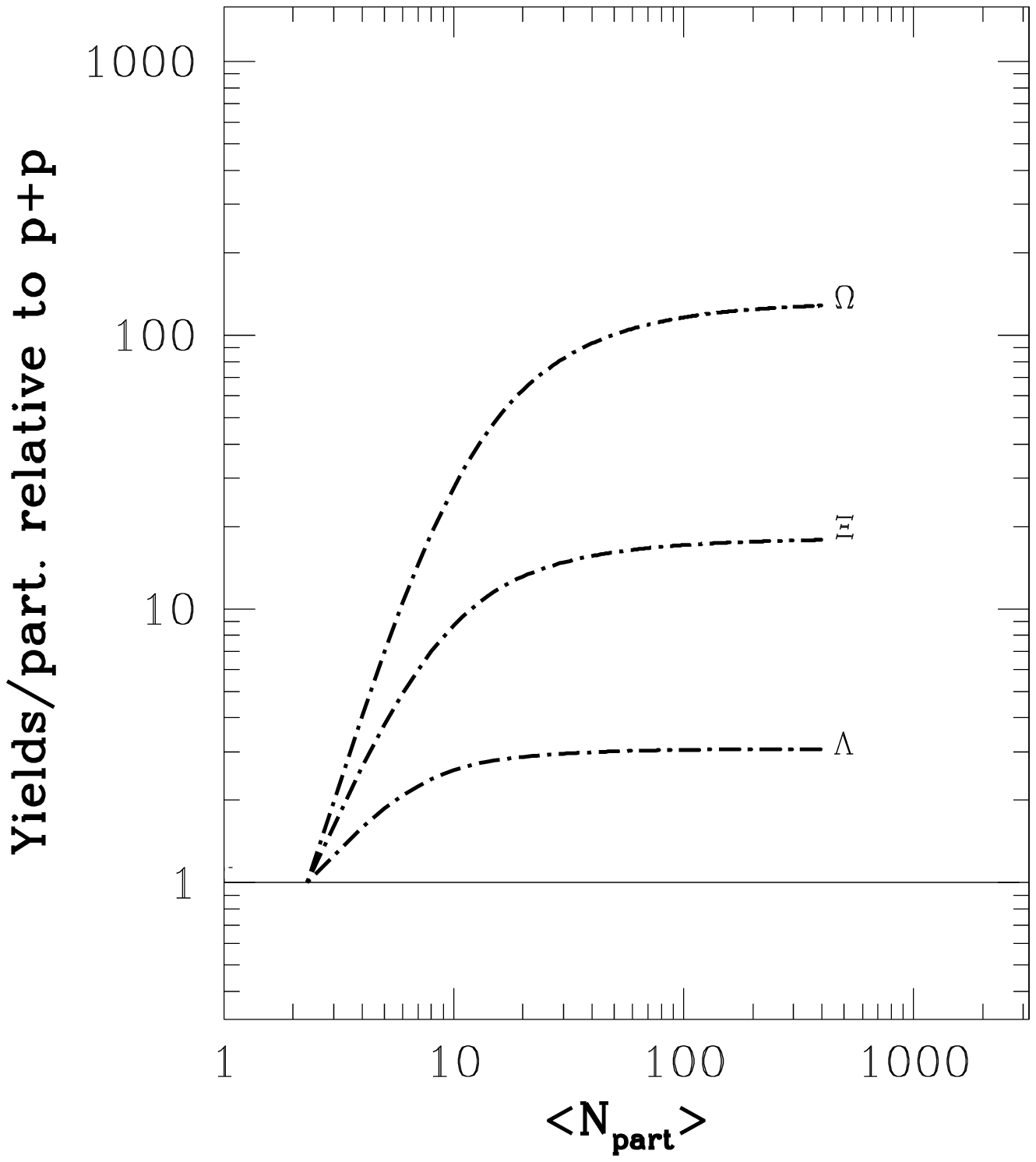} \vskip -1.cm
\caption{Centrality dependence of the relative enhancement of
particles in central Pb-Pb
 at  energy $\sqrt{s}=8.73$ GeV.\hfill}
\end{minipage}
\end{figure}
From Eq.(\ref{enh}), SE for a given $V\sim
N_{part}$ relative to pp is determined from
\begin{equation}
E_s^i= { F_s(T,V)\over F_s(T,V_{pp})}\equiv
{ F_s(T,N_{part})\over F_s(T,2)}
\end{equation}
Figure 1 displays an  example of the suppression factor at $\sqrt
s=8.73$ GeV.
 The enhancement hierarchy at corresponding energy, as seen in Figure 2, is a direct
consequence of the behavior of the suppression factor with the
strangeness content of the particle \cite{tounsi}. Thus,
SE can indeed appear as a canonical
suppression effect.
\section{ENERGY DEPENDENCE OF STRANGENESS ENHANCEMENT}
We have studied \cite{redlich} the energy dependence of SE at four
energies:$\sqrt s = 8.73, \ 12.3 , \ 17.3$  GeV  (NA49, NA57,
WA97, SPS) and $\sqrt s = 130$ GeV (RHIC). At 17.3   and 130 GeV
the values of  $T$ and $\mu_B$ were taken from the thermal fit
from \cite{braun,braun1}. At $\sqrt s=$ 8.73 and 12.3 GeV  the
values of these parameters were determined following   unified
freeze-out curve \cite{cleymans1} and by using the measured
$<\pi>/<N_{part}>$ ratio at the relevant energies. Moreover, as a
first approximation, the assumption have been made  that the
thermal parameters are independent of centrality. We have shown
that SE is a decreasing function of collision
energy \cite{redlich}. The result for $\Lambda$ is given in Figure
3. In Figure 4 the model  predictions are compared to the recent
experimental data of NA49 Collaboration \cite{mischke}. The
experimental points are deduced from the
$(<\Lambda>/<\pi>)^{AA}/(<\Lambda>/<\pi>)^{pp}$ ratio
\cite{mischke} multiplied by $ <\pi>/<N_{part}>$ in AA and in pp
collisions  \cite{marek} at the same energy. The solid line in
Figure 4 is the prediction of  the canonical statistical  model.

\begin{figure}[htb]
\vskip -1.5cm
\begin{minipage}[t]{74mm}
\includegraphics[width=17.5pc, height=14.5pc]{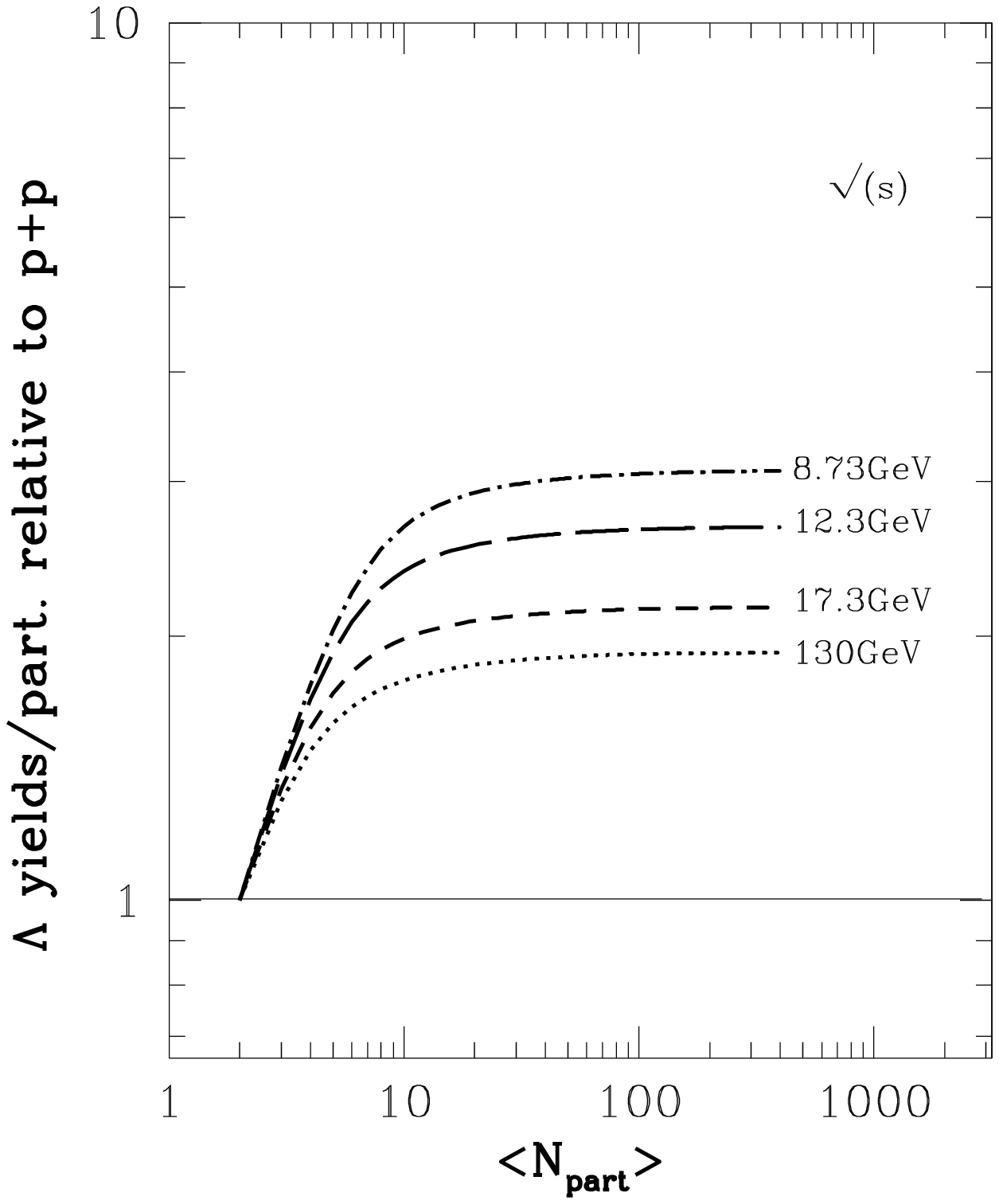}
\vskip -1.cm \caption{Centrality dependence of the relative
enhancement of $\Lambda$ in central Pb-Pb
 at  different collision energies.\hfill}
\end{minipage}
\hspace{\fill}
\begin{minipage}[t]{74mm}
\vskip -6.cm\hskip 1cm
\includegraphics[angle=180,width=12.pc]{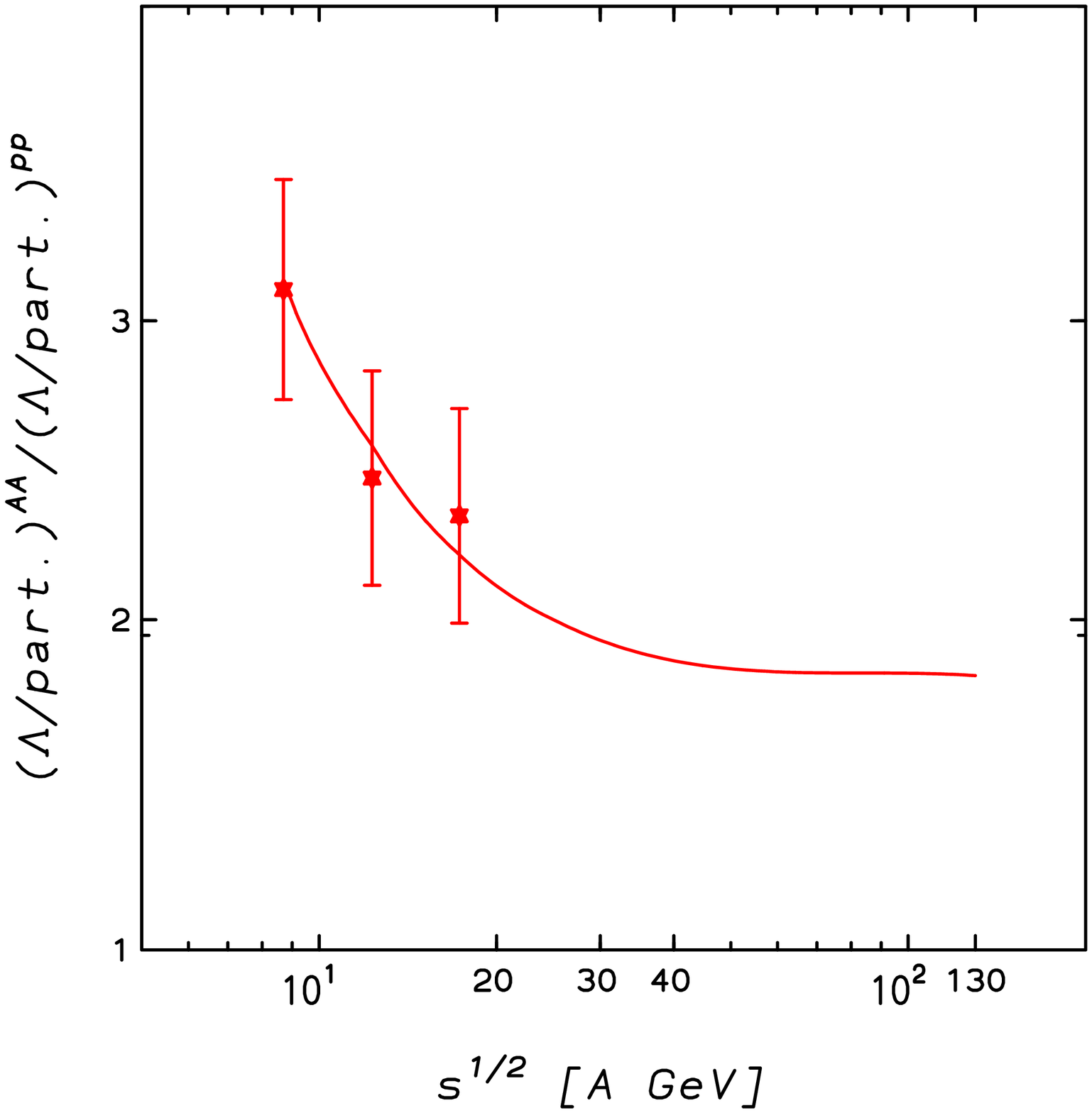}
\vskip -1.6cm \caption{Energy dependence of the relative
enhancement of $\Lambda$ in central Pb-Pb collisions.\hfill}
\end{minipage}
\end{figure}
\section{CONCLUDING REMARKS}
Canonical suppression  plays an
important role in SE. The present canonical
statistical model naturally describes \cite{hamieh} the bulk
properties of the WA97 data for $\Lambda, \Xi, \Omega$, that show
SE  hierarchy, and {\it saturation} of the
yields per participant for large $N_{part}$. However, the
quantitative understanding of  the recent data of the NA57
Collaboration \cite{manzari}, showing a smooth increase of that
yields with $N_{part}>100$ and a significant change in the
behavior of $\bar{\Xi}$ ( the yield rises from $N_{part}=62$ to
$N_{part}=121$ by a factor 2.6) would required a further study  on
the behavior of thermal parameters with centrality
\cite{cleyman2}.  These new features of NA57 data may be
understood admitting a non-linear dependence of the canonical
correlation volume or of the off-equilibrium strangeness fugacity
with centrality \cite{new}. Nevertheless, the enhancement
hierarchy and the decrease of enhancement with increasing
collision energy are a generic features of the canonical
statistical thermal model that are independent on the particular
choice of the parameters. The most recent  NA49 data on $\Lambda$
yields seem to support the model prediction that SE
 is a decreasing function of collisions  energy.
However, experimental confirmation of these behavior by even more
spectacular variation of $\Xi$ and $\Omega$ enhancement with
$\sqrt s$  would be required. We note that UrQMD \cite{bleicher}
and DPM \cite{capella} predicts a larger SE
at RHIC than at SPS energy, which is in contrast to present
statistical model results.

{\bf Acknowledgement:} We wish to thank P. Braun-Munzinger, J.
Cleymans, M. Gazdzicki,  H. Oeschler, H. Satz and R. Stock  for interesting
discussions. K.R acknowledges the support of the Alexander von
Humboldt Foundation.

\end{document}